\documentclass[prl,amssymb,twocolumn,superscriptaddress,showpacs]{revtex4}

\usepackage{bm}
\usepackage{graphicx}
\usepackage{color}
\usepackage{amsmath, amssymb, mathrsfs}
\usepackage[amssymb]{SIunits}
\usepackage{rotating}
\usepackage{pifont}
\usepackage{amsmath,amssymb,natbib,bm,graphicx,url,epsfig,color,verbatim,psfrag}
\usepackage[ansinew]{inputenc}

\begin{document}

\title{Cotunneling through a magnetic single-molecule transistor based on $\rm{N@C_{60}}$}

\author{Nicolas Roch}
\author{Romain Vincent}
\affiliation{Institut N\'eel, associ\'e \`a l'UJF, CNRS, BP 166,
38042 Grenoble Cedex 9, France}
\author{Florian Elste}
\affiliation{Department of Physics, Columbia University, 538 West
120th Street, New York, NY 10027, USA}
\author{Wolfgang Harneit}
\affiliation{Institut f\"ur Experimentalphysik, Freie Universit\"at
Berlin, Arnimallee 14, 14195 Berlin, Germany}
\author{Wolfgang Wernsdorfer}
\affiliation{Institut N\'eel, associ\'e \`a l'UJF, CNRS, BP 166,
38042 Grenoble Cedex 9, France}
\author{Carsten Timm}
\affiliation{Institut f\"ur Theoretische Physik, Technische Universit\"at Dresden,
 01062 Dresden, Germany}
\author{Franck Balestro}
\affiliation{Institut N\'eel, associ\'e \`a l'UJF, CNRS, BP 166,
38042 Grenoble Cedex 9, France}

\date{\today}
\begin{abstract}
We present an experimental and theoretical
study of a magnetic single-molecule
transistor based on $\rm{N@C_{60}}$ connected to gold electrodes. Particular attention is paid to
the regime of intermediate molecule-lead coupling, where
cotunneling effects manifest themselves in the Coulomb-blockade regime.
The experimental results for the differential conductance as
a function of bias, gate voltage, and external magnetic field are in agreement with our analysis of the tunneling rates and provide evidence of magnetic
signatures in single-$\rm{N@C_{60}}$ devices arising from an antiferromagnetic
exchange interaction between the $\rm{C_{60}}$ spin and the nitrogen spin.
\end{abstract}

\maketitle

\begin{figure}
\includegraphics[scale=0.2]{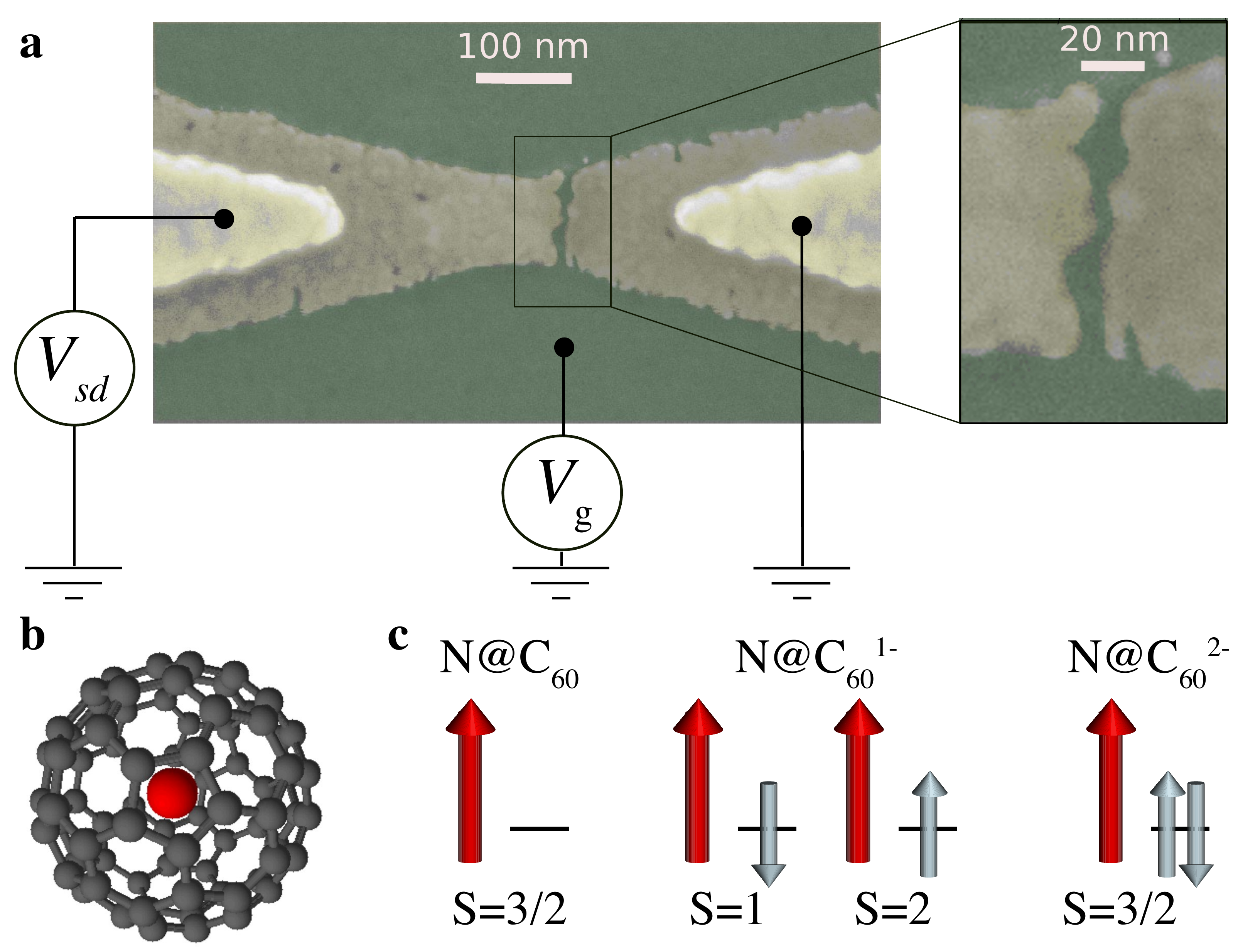}
\caption{(a) Colorized electron beam
microscopy photography of an SMT consisting of a gold nanowire over an
$\rm{Al/Al_{2}O_{3}}$ gate after electromigration. (b)
Schematic picture of $\rm{N@C_{60}}$.
(c) Spins states of neutral and charged $\rm{N@C_{60}}$.}
\label{figure1}
\end{figure}

Single molecules carrying a magnetic moment, such as $\rm{Mn_{12}}$, $\rm{Fe_{8}}$ and
$\rm{N@C_{60}}$, have been studied intensively for their outstanding quantum
properties. Some of these molecules are characterized by a large
spin and exhibit macroscopic quantum
tunneling~\cite{Friedman1996,Thomas1996} and quantum-interference effects~\cite{Wernsdorfer1999}. The ability to initialize and
manipulate the quantum state of a single magnetic molecule
could open the road to new strategies for high-density information
storage, quantum computing and molecular
spintronics~\cite{Harneit2002,Wernsdorfer2008}. However, only few experiments
have succeeded in fabricating single-molecule
transistors (SMTs) [Fig.~\ref{figure1}(a)] based on magnetic
molecules~\cite{Heersche2006,Jo2006,Henderson2007,Grose2008,Zyazin2010}. Furthermore, only two 
experiments~\cite{Jo2006,Zyazin2010} on single magnetic molecules
in three-terminal geometry have succeeded to obtain the intermediate molecule-lead coupling 
regime. In this regime~\cite{Moth2009},
cotunneling processes manifest
themselves in the spectra of excited molecular states for a fixed 
charge state of the molecule, allowing an easier and more 
precise characterization of magnetic properties of 
single-molecule devices compared to sequential tunneling.
For Mn$_{12}$ clusters,
steps in $dI/dV$ have been observed~\cite{Jo2006}, but
the dependence of the cotunneling steps on an external magnetic field 
was not investigated. Recently, a characteristic zero-field 
splitting and its evolution in an external magnetic field has been
attributed to the signature of anisotropy in a $\rm{Fe_{4}}$ 
single-molecule magnet~\cite{Zyazin2010}. The aim of our work
is then to study exchange-coupling-dependent cotunneling features in 
a single magnetic molecule.

Previous
studies have shown that it is not straightforward to
identify the magnetic properties of a SMT since molecular magnetism is
usually destroyed during the device fabrication~\cite{Jo2006}
due to strong interactions with the environment~\cite{Manini2008}. To conserve
the magnetic properties, one option would be to incorporate a molecule whose
magnetic moment is retained.
In $\rm{N@C_{60}}$, the spin of the encapsulated nitrogen atom,
which sits in the center of the fullerene molecule~\cite{Mauser1997},
is protected by the $\rm{C_{60}}$ cage~[Fig.~\ref{figure1}(b)].
An exchange interaction
between the $\rm{C_{60}}$ spin and the nitrogen spin
permits an indirect measurement of
the magnetic properties of the nitrogen~\cite{Grose2008}.

Although experimental studies investigating the stability of paramagnetism in $\rm{N@C_{60}}$ on a substrate have not been performed so far,
it is known from X-ray and ultraviolet photoelectron spectroscopy that
$\rm{C_{60}}$
changes its charge state when being adsorbed on noble metal surfaces~\cite{Modesti1993,Swami1999}.
This may lead to the formation of $\rm{N@C_{60}^{1-}}$ ions, when the number of
electrons, $n_{\rm{C_{60}}}$, added to the $\rm{C_{60}}$ cage is 1, or $\rm{N@C_{60}^{2-}}$ ions when $n_{\rm{C_{60}}}=2$.
Since covalently functionalized
$\rm{N@C_{60}}$ is
stable~\cite{Dietel1999,Goedde2001,Franco2006}, various ions of $\rm{N@C_{60}}$
are believed to be stable~\cite{Jakes2000}. 

The total spin of neutral
$\rm{N@C_{60}}$ is $S= 3/2$ [Fig.~\ref{figure1}.(c)].
However, the spin of the singly-charged ion $\rm{N@C_{60}^{1-}}$
can be $S=1$ or $S=2$, depending on whether
the $\rm{C_{60}}$ spin $S_{\rm{C_{60}}}= 1/2$ is
antiparallel or parallel
to the nitrogen spin
[Fig.~\ref{figure1}(c)].
$\rm{N@C_{60}^{2-}}$ has a fixed ground state spin $S= 3/2$ since the 
dianion cage is diamagnetic due to Jahn-Teller distortion.

Following theoretical predictions~\cite{Elste2005},
transport measurements
on SMTs with $\rm{N@C_{60}}$
in the weak-coupling regime showed
signatures of an antiferromagnetic exchange
interaction between the $\rm{C_{60}}$ spin and the nitrogen spin 
\cite{Grose2008}, in contradiction with the 
ferromagnetic exchange coupling obtained, for singly and triply ionized anions of 
$\rm{N@C_{60}}$, using 
multi-configurational self-consistent field 
methods~\cite{Udvardi2000}. Therefore, we first present
results that confirm
the interpretation of previous experiment~\cite{Grose2008} in the weak-coupling regime.
As the field of molecular electronics in general and of molecular
spintronics in particular has been plagued by a lack of 
reproducibility, it is essential that we have been able to reproduce the results for
the prototypical single-molecule transistor based on $\rm{N@C_{60}}$ using independently fabricated
devices.
However, since we succeeded in reaching the intermediate molecule-lead coupling
regime, our results go
beyond the sequential-tunneling regime. In the second part,
we present results in the cotunneling regime.
In agreement with our theoretical
analysis, our experimental results provide evidence for an
antiferromagnetic exchange interaction between the $\rm{C_{60}}$ spin 
and the nitrogen spin.

The SMT was fabricated by using the electromigration
technique~\cite{McEuen1999}.
The experiment was carried out in a dilution
refrigerator at a base temperature of $35\,\mathrm{mK}$.
Our procedure and details of the measurement
system are described in Ref.~\cite{RochJLTP}.

\begin{figure}
\includegraphics[scale=0.46]{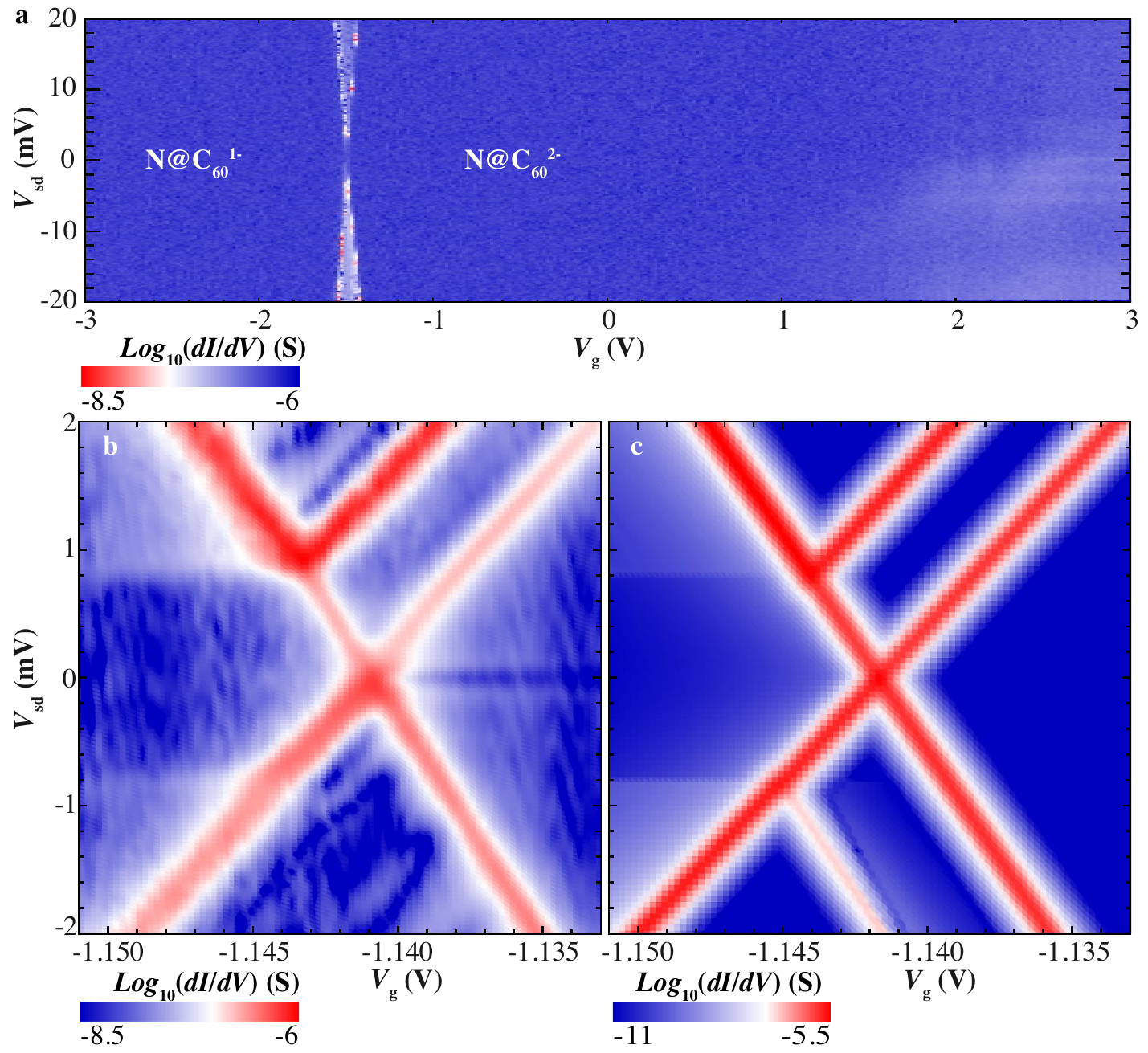}
\caption{Color scale plot of the differential conductance
$dI/dV$ as a function of bias voltage $V_{\rm{sd}}$ and gate voltage
$V_{\rm{g}}$ (a) on the entire accessible $V_{\rm{g}}$ range and (b)
in the vicinity of $V_{\rm{g}}^{\rm{D}}$. (c) Numerical results
obtained from Eqs.~(\ref{seq-rates}) and
(\ref{cot-rates}) for $\varepsilon_d = -3.1625 \,\mathrm{eV}$,
$U = 3.0 \,\mathrm{eV}$, $J= -0.4 \,\mathrm{meV}$, $C_{g}$ : $C_{R}$ : $C_{L}$ = 1 : 3.09 : 2.96,
and $\Gamma_R/\Gamma_L\simeq
10$.}
\label{figure2}
\end{figure}

Figure~\ref{figure2}(a) shows a color scale plot
of the differential conductance
$dI/dV$ as a function of bias voltage $V_{\rm{sd}}$ and
gate voltage $V_{\rm{g}}$ on the entire accessible gate range, performed
during a first run of measurements.
We emphasize that we observe only
one charge degeneracy point (at gate voltage $V_{\rm{g}}^{\rm{D}}$), where
the energy of the $n_{C_{60}}=1$ state aligns with energy of the
$n_{C_{60}}=2$ state.
Figure~\ref{figure2}(b) shows results for $dI/dV$ in the vicinity of
$V_g^D$.
Analyzing the slopes of the stability
diagram, we obtain $C_{g}$ : $C_{R}$ : $C_{L}$ = 1 :
3.09 : 2.96 for the ratio of the capacitances, where $C_{g}$, $C_{R}$ and $C_{L}$ denote the
capacitances of the gate, the source, and the drain with respect to
the molecule. Thus the effective gate voltage is reduced by a factor of
$C_{g}/(C_{g}+C_{R}+C_{L}) = 0.142$. From this we infer that the
energy spacing between the relevant charge degeneracy points
is at least larger than $230\,\mathrm{meV}$ and conjecture that the
measurement was performed for a single $\rm{N@C_{60}}$ in the junction~\cite{Ward2008}.

When the energy level spacing between the molecular states is
large compared to the thermal energy,
the hybridization energies $\Gamma_{R}$ and $\Gamma_{L}$ can be estimated
by an analysis of the current amplitudes
at different gate voltages~\cite{Bonet2002}.
We find $\Gamma_L \simeq 1.65\, \mu \rm{eV}$ and $\Gamma_R \simeq 16.5 \, \mu \rm{eV}$
($\Gamma_R/\Gamma_L \simeq 10$).
The assumption of asymmetric
molecule-lead couplings explains the different peak heights
in $dI/dV$ at negative and positive bias [Fig.~\ref{figure2}(b)] and is taken into
account in the theoretical treatment discussed below [Fig.~\ref{figure2}(c)].

We compare the bias- and gate-voltage dependences of the
measured differential conductance to numerical results obtained from a rate-equation approach.
The system is described
by a Hamiltonian of the form $H = H_{\mathrm{N@C_{60}}} + H_{\text{leads}} + H_t$,
where
\begin{align}\label{H_NC60}
H_{\mathrm{N@C_{60}}} ~=~ &
 \varepsilon_d \, n_\mathrm{C_{60}} + \frac{U}{2} n_\mathrm{C_{60}} (n_\mathrm{C_{60}}-1) \nonumber \\
 & - J\, \mathbf{S}_{\mathrm{C_{60}}} \cdot \mathbf{S}_{\mathrm{N}}
 - g\mu_B B \left( S^z_{\mathrm{C_{60}}} + S^z_{\mathrm{N}} \right)
\end{align}
describes the molecular degrees of freedom,
\begin{equation}
H_{\text{leads}} = \sum_{\alpha=L,R} \sum_{\mathbf{k} \sigma}
  \epsilon_{\alpha \mathbf{k}}
  a_{\alpha \mathbf{k} \sigma}^{\dagger} a_{\alpha \mathbf{k} \sigma}
\end{equation}
describes the conduction electrons in the leads, and
\begin{equation}\label{H_t}
H_t= \sum_{\alpha=L,R} \sum_{\mathbf{k} \sigma}
  (t_{\alpha} a_{\alpha \mathbf{k} \sigma}^{\dagger} d_{\sigma}
  + t_{\alpha}^{\ast} d_{\sigma}^{\dagger} a_{\alpha \mathbf{k} \sigma})
\end{equation}
describes the tunneling of electrons between the $\mathrm{C_{60}}$
cage and the leads.
Here, $d_{\sigma}^{\dagger}$ creates an electron with spin $\sigma$
and energy $\varepsilon_d$ in the LUMO of the
$\mathrm{C_{60}}$ molecule, while $a_{\alpha
\mathbf{k} \sigma}^{\dagger}$
creates an electron in lead $\alpha$
with spin $\sigma$, momentum $\mathbf{k}$, and
energy $\epsilon_{\alpha \mathbf{k}}$.
Tuning the gate voltage $V_{\rm{g}}$ allows one to shift the orbital
energy $\varepsilon_d$.
The number and spin operator of the $\mathrm{C_{60}}$ electrons
are $n_\mathrm{C_{60}}=\sum_{\sigma} d_{\sigma}^\dagger d_{\sigma}$
and $\textbf{S}_{\mathrm{C_{60}}}=\sum_{\sigma\sigma'} d_{\sigma}^\dagger
(\mbox{\boldmath$\sigma$}_{\sigma\sigma'}/2) d_{\sigma'}$, respectively, with
$\mbox{\boldmath$\sigma$}$ the vector of the Pauli matrices.
The exchange interaction with the nitrogen spin $\textbf{S}_{\mathrm{N}}$ is denoted by $J$ and
the external magnetic field applied along the $z$ axis by
$B$, where $g$ is the $g$-factor and $\mu_{B}$ is the Bohr
magneton.

The leading contribution to the tunneling rates (corresponding to sequential tunneling)
is of second order in $H_t$ and can be obtained from Fermi's Golden Rule,
\begin{align}\label{seq-rates}
R^{\text{(seq)}}_{n\rightarrow n',\alpha} &~=~ \Gamma_\alpha \sum_\sigma \Big( f(\epsilon_{n'}-\epsilon_n-\mu_{\alpha})|C^\sigma_{nn'}|^2 \nonumber \\
&~~~~~~ + \left[1-f(\epsilon_n-\epsilon_{n'}-\mu_{\alpha})\right]
|C^\sigma_{n'n}|^2 \Big).
\end{align}
Here $n$ and $n'$ denote
the initial and final many-particle state of $\mathrm{N@C_{60}}$, and
$f$ is the Fermi function.
The matrix element with $C^{\sigma}_{nn'}\equiv
\langle n | d_{\sigma} | n' \rangle$
denotes the net overlap of the initial and final states of the molecule. The typical
sequential-tunneling rate is given by $\Gamma_{\alpha}=2\pi t_\alpha2 \nu_\alpha$,
where $\nu_\alpha$ denotes the lead density of states ($\hbar=1$).
To mimick the finite life-time broadening of the sequential-tunneling peaks, we used an
effective temperature of $250\,\mathrm{mK}$ in Eq.\ (\ref{seq-rates}).

The next-to-leading-order contribution to the tunneling rates
(including cotunneling) is of fourth order in $H_t$ and can be computed using
a $T$-matrix formalism \cite{Bruus,Elste}.
We find
\begin{align}
R^{\text{(cot)}}_{n\rightarrow n',\alpha \rightarrow \alpha'} &= \frac{\Gamma_\alpha \Gamma_{\alpha'}}{2\pi} \,
  \sum_{\sigma \sigma'} \int d\epsilon \bigg\{ \bigg|
  \sum_{n''} \frac{ C^{\sigma \ast}_{n''n'} C^{\sigma'} _{n''n}}
  {\epsilon_{n'}-\epsilon_{n''}-\epsilon+i\eta} \bigg|^2 \nonumber \\
~ & {}+ \bigg| \sum_{n''} \frac{ C^\sigma _{n'n''} C^{\sigma' \ast}_{nn''}}
  {\epsilon_{n}-\epsilon_{n''}+\epsilon+i\eta} \bigg|^2 \bigg\} \nonumber \\
~~~~ \times & f(\epsilon-\mu_\alpha)\,
  \big[ 1-f(\epsilon_{n} - \epsilon_{n'} + \epsilon-\mu_{\alpha'})\big]. \label{cot-rates}
\end{align}
The divergences arising from the energy denominators are regularized by replacing $\eta$ by
a finite life-time broadening $\Gamma=\Gamma_L+\Gamma_R$. In the Coulomb-blockade regime
far from the sequential-tunneling threshold, the Fermi functions suppress
the integrand in Eq.\ (\ref{cot-rates}) outside of a narrow energy interval.
Replacing $\epsilon$ in the energy denominators by its mean on these intervals,
we can perform the integral. Close to the sequential-tunneling threshold, where this
approximation becomes invalid, the cotunneling contribution is overwhelmed by the
sequential tunneling in any case.

The good qualitative agreement between
Figs.~\ref{figure2}(b) and \ref{figure2}(c) suggests that
the Hamiltonian [Eqs.~(\ref{H_NC60})--(\ref{H_t})] contains the
crucial ingredients to explain both sequential and cotunneling features revealed in the experiment.
We find that a calculation under the assumption of $J \simeq -0.4 \,\mathrm{meV}$
gives the best agreement with the experiment.

\begin{figure}
%\begin{center}
\includegraphics[scale=0.65]{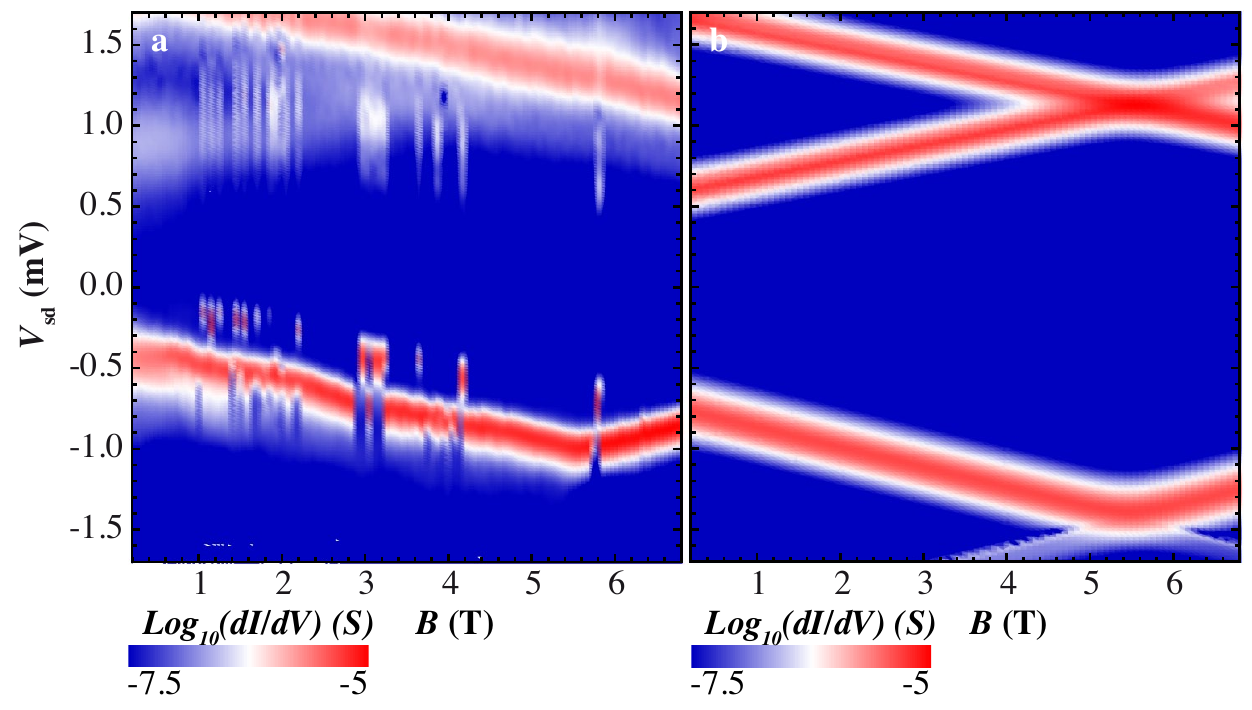}
%\end{center}
\caption{Color scale plot of the differential conductance
$dI/dV$ as a function of bias voltage and magnetic field $B$.
(a) shows experimental results while (b) shows numerical results in the regime
where the ground state is $n_\mathrm{C_{60}} = 2$, with $\varepsilon_d = -3.1722 \,\mathrm{eV}$,
$U = 3.0 \,\mathrm{eV}$, $J= -0.3 \,\mathrm{meV}$,
$C_{g}$ : $C_{R}$ : $C_{L}$ = 1 : 3.29 : 3.43,
and $\Gamma_R/\Gamma_L\simeq 6.7$.}
\label{figure3}
\end{figure}

In order to verify that the charge degeneracy point involves the two states
$n_\mathrm{C_{60}}=1$ and $n_\mathrm{C_{60}}=2$, we investigated the
dependence of $dI/dV$ on an external magnetic field $B$.
These measurements were performed for the same sample but during a different run.
Results of $dI/dV$ for a fixed gate voltage greater
than $V_{\rm{g}}^{\rm{D}}$ are shown in Fig.~\ref{figure3}.
The conductance peaks at positive and negative
bias voltage first move apart until the magnetic field assumes a critical value.
Then they change slope approaching each other again. Here the best fit to
experimental data is obtained for a calculation
that assumes $J \simeq -0.3 \,\mathrm{meV}$.
The other parameters are also changed somewhat, see the caption of Fig.~\ref{figure3}.
The microscopic mechanism behind the change in the exchange interaction is not yet clear.
The characteristic pattern revealed in Fig.~\ref{figure3}
allows us to deduce the charge and spin of the relevant molecular states to the
left and to the right of the degeneracy point, in agreement with the results reported
by Grose {\it et al.}~\cite{Grose2008}.

%\section{Cotunneling regime}

We now turn to the cotunneling regime where transport is dominated by
the coherent tunneling of electrons through the source-molecule-drain structure~\cite{Averin1992,Franceschi2001}. This regime
is characterized by a fixed molecular charge state and by the occurrence of
steps in $dI/dV$. The positions of these steps are independent of
the molecular charging energies and the gate voltage but do depend
on the magnetic excitation energies.
As a consequence, measurements in this regime can be more easily interpreted than
in the sequential-tunneling regime, where two different charge states
are involved.
%This allows for a precise magnetic spectroscopy that does not require the
%deconvolution of the energy spectra
%of two different charge states.

\begin{figure}
%\begin{center}
\includegraphics[scale=0.65]{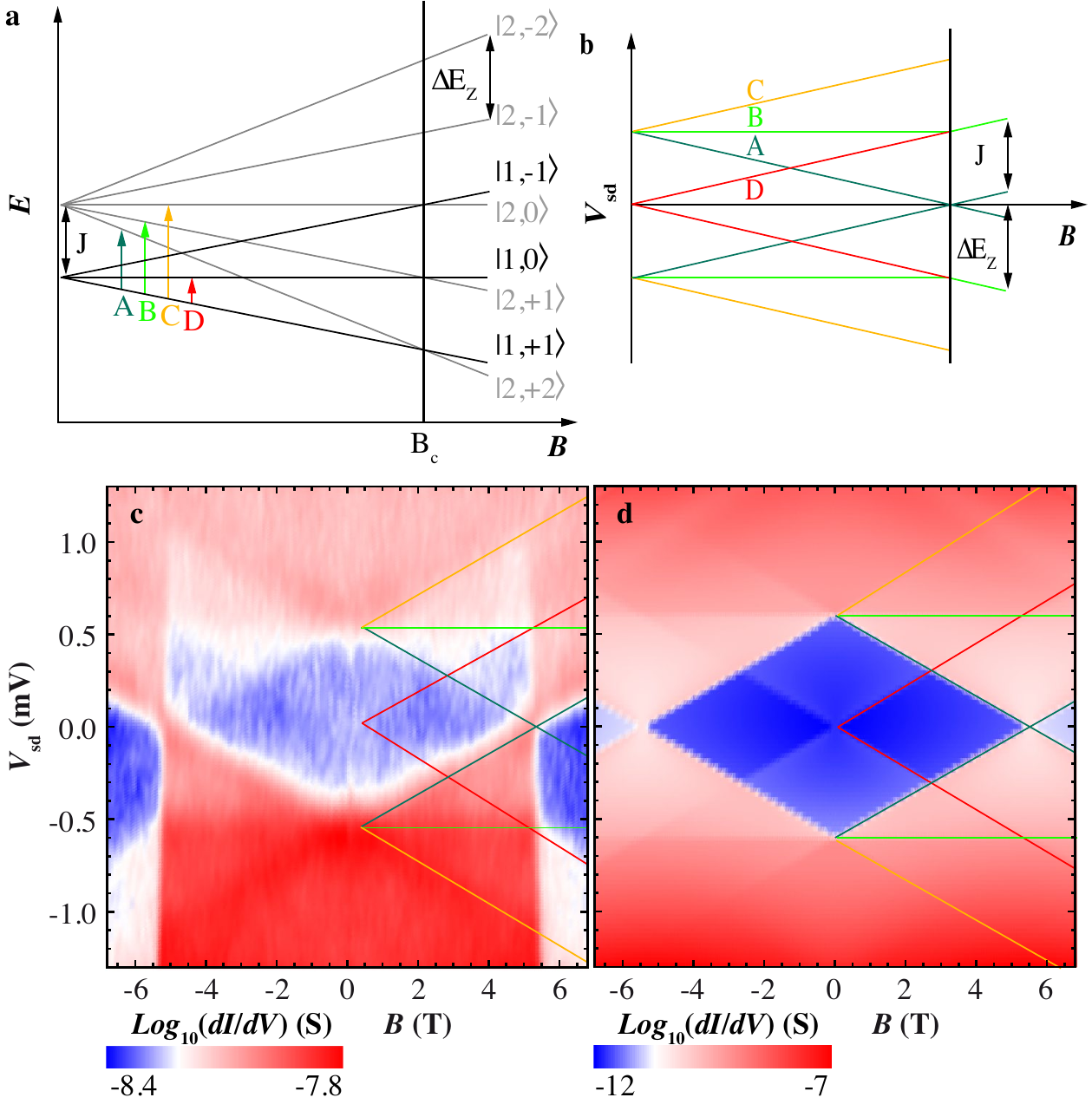}
%\end{center}
\caption{(a) Zeeman diagram of the multiplets $S = 1$ and $S
= 2$ for $\rm{N@C_{60}^{1-}}$ ($n_\mathrm{C_{60}}=1$). (b) Schematic showing the expected
cotunneling steps. (c) Experimental and (d) numerical results,
with  $J= -0.3 \,\mathrm{meV}$, of $dI/dV$
for a fixed gate voltage $V_{\rm{g}} < V_{\rm{g}}^{\rm{D}}$. Colored
lines indicate the positions of the magnetic excitations illustrated in (a) and (b).}
\label{figure4}
\end{figure}

For $\rm{N@C_{60}^{1-}}$, the ground
state is a degenerate spin $S=1$ multiplet
while the excited state, having an energy $J$ compared to the ground 
state, is a spin $S = 2$ multiplet at zero magnetic 
field. Turning on a magnetic field
leads to a Zeeman splitting of the spin multiplets as indicated in
Fig.~\ref{figure4}(a). Selection rules for the cotunneling processes
require $\left|\Delta S^{z}\right| = 0 \; \rm{or} \; 1$. This condition is used to determine all possible
transitions that we expect to observe as steps in $dI/dV$ 
(Fig.~\ref{figure4}(b)).
Our measurement presented in Fig.~\ref{figure4}(c) is in qualitative 
agreement with our theoretical analysis presented 
in Fig.~\ref{figure4}(d), obtained with the same exchange coupling $J= -0.3 
\,\mathrm{meV}$, confirming in an another tunneling process, the antiferromagnetic exchange 
coupling between the $\rm{C_{60}}$ spin 
and the nitrogen spin.

The discussion for the case $V_g > V_g^D$,
where the ground state is a doubly-charged $\rm{N@C_{60}^{2-}}$ 
anion, proceeds in an analogous way and is included in the 
supplementary information.
The most striking difference is the presence of a cotunneling step at very
low bias, for zero magnetic field~[Fig.~\ref{figure2}(b)].
This feature might be due to magnetic anisotropy of the molecular
spin~\cite{Zyazin2010},
caused by the spin-orbit coupling of the nitrogen, which we expect to be enhanced
by the hybridization with the gold leads.

In summary, we have presented an experimental and
theoretical analysis of an SMT based
on $\rm{N@C_{60}}$. Our results in the sequential-tunneling
regime are in agreement with the findings previously reported~\cite{Grose2008}, 
and are a step towards achieving the high degree of reproducibility
necessary for progress in the field of molecular spintronics.
Furthermore, we have investigated the cotunneling regime, where
transport is dominated by the coherent transfer of electrons through
the molecular junction.
Since measurements in this regime can be more easily interpreted than
in the sequential-tunneling regime, the characteristic two-dimensional plots of
$dI/dV$ as a function of bias voltage and magnetic field
serve a characteristic fingerprint of the magnetic state.
In particular, co-tunneling data obtained
as a function of magnetic field corroborate the
antiferromagnetic exchange interaction between $\rm{C_{60}}$ spin and the nitrogen spin.

\begin{acknowledgments}

We thank E. Eyraud, D. Lepoittevin for technical
support, T. Crozes, T. Fournier for help with the lithographic methods,
S. Florens, V. Bouchiat, C. Winkelmann, L. Udvardi for fruitful discussions,
and C. Thirion,  E. Bonet, R. Piquerel for help with the software development.
Samples were fabricated in the NANOFAB
facility of the N\'eel Institute. This work is financially supported by
ANR-PNANO project  MolNanoSpin
n$^\circ$ANR-08-NANO-002, ERC Advanced Grant MolNanoSpin n$^\circ$226558, STEP
MolSpinQIP, Cible 2009, and the Deutsche Forschungsgemeinschaft.

\end{acknowledgments}

%%---------Biblio-------------%%

%\bibliographystyle{apsrev}
%\bibliography{biblio}

\begin{thebibliography}{10}






\bibitem{Friedman1996}
J. R. Friedman, M. P. Sarachik, J. Tejada, and R. Ziolo,
Phys. Rev. Lett. {\bf 76}, 3830 (1996).

\bibitem{Thomas1996}
L. Thomas {\it{et al.}},
Nature (London) {\bf 383}, 145 (1996).

\bibitem{Wernsdorfer1999}
W. Wernsdorfer and R. Sessoli,
Science {\bf 284}, 133 (1999).

\bibitem{Harneit2002}
W. Harneit,
Phys. Rev. A {\bf 65}, 032322 (2002).

\bibitem{Wernsdorfer2008}
L. Bogani and W. Wernsdorfer,
Nature Materials {\bf 7}, 179 (2008).

\bibitem{Heersche2006}
H. B. Heersche {\it{et al.}},
Phys. Rev. Lett. {\bf 96}, 206801 (2006).

\bibitem{Jo2006}
M. H. Jo {\it{et al.}},
Nanoletters {\bf 6}, 2014 (2006).

\bibitem{Henderson2007}
J. J. Henderson, C. M. Ramsey, E. D. Barco, A. Mishra, and
 G. Christou,
J. Appl. Phys. {\bf 101}, 09E102 (2007).

\bibitem{Grose2008}
J. E. Grose {\it{et al.}},
Nature Materials {\bf 7}, 884 (2008).

\bibitem{Zyazin2010}
A. S. Zyazin {\it{et al.}},
Nanoletters {\bf 10}, 3307 (2010).

\bibitem{Moth2009}
K. Moth-Poulsen and T. Bj{\o}rnholm,
Nature Nanotechnology {\bf 4}, 551 (2009).

\bibitem{Manini2008}
M. Mannini {\it et al}.
Chemistry-A European Journal, {\bf 14}, 7530 (2008).

\bibitem{Mauser1997}
H. Mauser {\it et al}.,
Angew. Chem. Int'l. Ed. {\bf 36}, 2835 (1997).

\bibitem{Modesti1993}
S. Modesti, S. Cerasari, and P. Rudolf,
Phys. Rev. Lett. {\bf 71}, 2469 (1993).

\bibitem{Swami1999}
N. Swami, H. He, and B. E. Koel,
Phys. Rev. B {\bf 59}, 8283 (1999).

\bibitem{Dietel1999}
E. Dietel {\it{et al.}},
J. Am. Chem. Soc. {\bf 121}, 2432 (1999).

\bibitem{Goedde2001}
B. Goedde {\it{et al.}},
Chem. Phys. Lett. {\bf 334}, 12 (2001).

\bibitem{Franco2006}
L. Franco {\it{et al.}},
Chem. Phys. Lett. {\bf 422}, 100 (2006).

\bibitem{Jakes2000}
P. Jakes {\it{et al.}}, in {\it Electronic Properties of
Novel-Materials-Molecular Nanostructures}, edited by H. Kuzmany {\it
et al.}, AIP Conf. Proc. {\bf 544}, 174 (AIP, Melville, NY, 2000).

\bibitem{Elste2005}
F. Elste and C. Timm,
Phys. Rev. B {\bf 71}, 155403 (2005).

\bibitem{Udvardi2000}
L. Udvardi, in {\it Electronic Properties of 
Novel-Materials-Molecular Nanostructures}, edited by H. Kuzmany {\it 
et al.}, AIP Conf. Proc. {\bf 544}, 187 (AIP, Melville, NY, 2000).

\bibitem{McEuen1999}
H. Park, A. K. L. Lim, A. P. Alivisatos, J. Park, and P. L. McEuen,
Appl. Phys. Lett. {\bf 75}, 301 (1999).

\bibitem{RochJLTP}
N. Roch, S. Florens, V. Bouchiat, W. Wernsdorfer, and F. Balestro,
J. Low Temp. Phys. {\bf 153}, 350 (2008).

\bibitem{Ward2008}
D. R. Ward, G. D. Scott, Z. K. Keane, N. J. Halas, and D. Natelson,
J. Phys. : Cond. Matter {\bf 20}, 374118 (2008).

\bibitem{Bonet2002}
E. Bonet, M. M. Deshmukh, and D. C. Ralph,
Phys. Rev. B {\bf 65}, 45317 (2002).

\bibitem{Bruus}
H. Bruus and K. Flensberg,
in {\it Many-body Quantum Theory in
Condensed Matter Physics},
(Oxford University Press, Oxford, 2004).

\bibitem{Elste}
F. Elste and C. Timm,
Phys.\ Rev.\ B \textbf{75}, 195341 (2007).

\bibitem{Averin1992}
D. V. Averin and Y. V. Nazarov, in {\it Single Charge Tunneling: Coulomb Blockade phenomena in
Nanostructures}, edited by H. Grabert and M. H. Devoret (Plenum Press
and NATO Scientific Affairs Division, New York, 1992) p 217.

\bibitem{Franceschi2001}
S. De Franceschi {\it{et al.}},
Phys. Rev. Lett. {\bf 86}, 878 (2001).

%\bibitem{Lips2000}
%K. Lips, M. Waiblinger, B. Pietzak, and A. Weidinger,
%Phys. Status Solidi A {\bf 177}, (2000)

%\bibitem{Naydenov2006}
%B. Naydenov {\it{et al.}},
%Chem. Phys. Lett. {\bf 424}, 327 (2006)



%\bibitem{Ave94}
%D. V. Averin,
%Physica B \textbf{194-196}, 979 (1994).

%\bibitem{TuM02}
%M. Turek and K. A. Matveev, Phys.\ Rev.\ B \textbf{65}, 115332 (2002).



\end{thebibliography}

\end{document}

%opening
\title{Supplementary information : Cotunneling through a magnetic
single-molecule transistor based on $\rm{N@C_{60}}$}

\author{Nicolas Roch}
\author{Romain Vincent}
\affiliation{Institut N\'eel, associ\'e \`a l'UJF, CNRS, BP 166,
38042 Grenoble Cedex 9, France}
\author{Florian Elste}
\affiliation{Department of Physics, Columbia University, 538 West 
120th Street, New York, NY 10027, USA}
\author{Wolfgang Harneit}
\affiliation{Institut f\"ur Experimentalphysik, Freie Universit\"at 
Berlin, Animallee 14, 14195 Berlin, Germany}
\author{Wolfgang Wernsdorfer}
\affiliation{Institut N\'eel, associ\'e \`a l'UJF, CNRS, BP 166,
38042 Grenoble Cedex 9, France}
\author{Carsten Timm}
\affiliation{Institut f\"ur Theoretische Physik, Technische
Universit\"at
Dresden,
 01062 Dresden, Germany}
\author{Franck Balestro}
\affiliation{Institut N\'eel, associ\'e \`a l'UJF, CNRS, BP 166,
38042 Grenoble Cedex 9, France}

\maketitle

\section{Samples presentation and comparison of the results}

\begin{figure}[h]
\includegraphics[width=16cm]{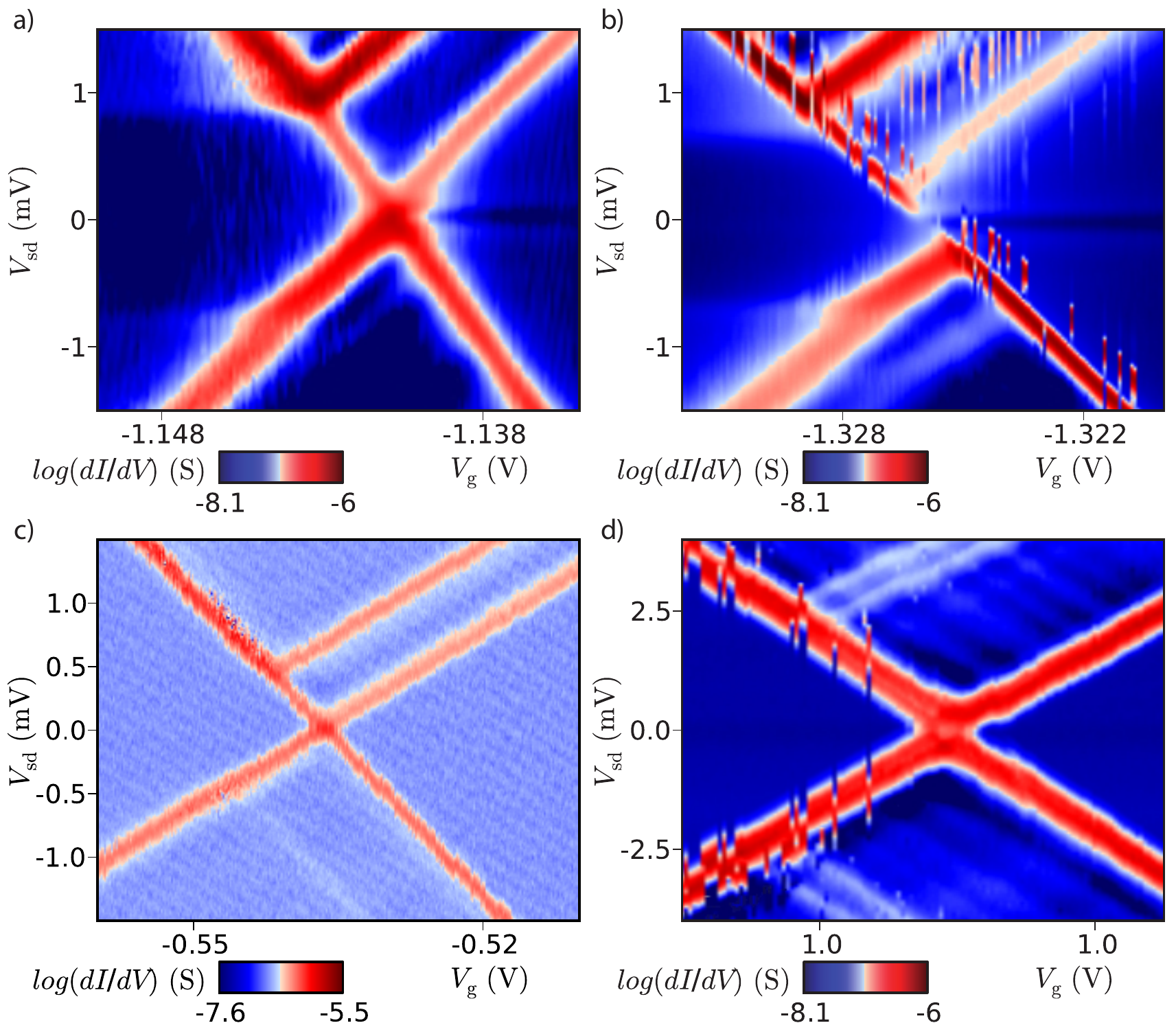}
\caption{{ Color scale plot of the differential conductance 
$dI/dV$ as a function of bias voltage $V_{\rm{sd}}$ and gate voltage 
$V_{\rm{g}}$} : {\bf a)} first run of measurements. {\bf b)} 
second run of measurements. {\bf c)} and {\bf d)} other 
$\rm{N@C_{60}}$ SMT's
in the weak coupling regime.}
\label{figure1}
\end{figure}

Single-molecule transistors (SMT's) based on 
$\rm{N@C_{60}}$ contacted by gold leads were prepared by blow drying 
a dilute toluene solution of the $\rm{N@C_{60}}$ molecule onto a gold 
nano-wire realized on an $\rm{Al/Al_{2}O_{3}}$ back gate. Before blow 
drying the solution, the nano-wires were cleaned with acetone, 
ethanol, isopropanol and oxygen plasma. We then performed our 
electromigration procedure at $4$~K. We tested 274 junctions and 
present in this section $3$ samples having an addition energy 
greater than $200$~meV~(Fig.S.\ref{figure1})

\def\figurename{Fig.S.}

In Fig.S.\ref{figure1}(a) and 1(b) we 
present first the stability diagram of the sample presented in the 
Letter, obtained at zero magnetic field during 
the first and the second run of measurements, respectively. 
The sample has been kept at cryogenic temperature ($T\leq 4$ K) during
the two
runs. However, changes in our electronic setup~(different I-V 
converters, inversion of the ground connection) have probably
generated 
a small electrical discharge inducing modifications of some
measurement features, as exhibited by the modification of the 
capacitances and tunneling rates extracted from a fine analysis of 
the stability diagram.
For the first run, we obtain $C_{g}$ : $C_{R}$ : $C_{L}$ = 1 :
3.09 : 2.96, and $\Gamma_R/\Gamma_L \simeq 10$,
while for the second run we get $C_{g}$ : $C_{R}$ : $C_{L}$ = 1 :
3.29 : 3.43, and $\Gamma_R/\Gamma_L \simeq 6.7$. The SMT experienced 
then some unexpected changes inducing a
different electrostatic environment leading to a modification of the
anti-ferromagnetic exchange interaction $J$. The evolution of $J$ will be addressed in the
next
section.

The striking difference between the two runs
is observed at the charge degeneracy point.
In Fig.S.\ref{figure1}(b), corresponding to the second run, we measure an opening of this latter. As
it is not present 
in the first run as seen in Fig.S.\ref{figure1}(a), we can rule out
any correlation between this observation and the cotunneling step at low
bias voltage
observed for $n_{\rm{C_{60}}}=2$ ($V_{\rm{g}} > V_{\rm{g}}^{\rm{D}}$). No further measurements were
performed in
order to explain this feature, but a similar behavior was observed in
another SMT
based on a single $\rm{N@C_{60}}$, as presented in
Fig.S.\ref{figure1}(d).

\section{Magnetic field spectroscopy in the
cotunneling
regime for 
$n_{\rm{C_{60}}}=1$}

\begin{figure}
\includegraphics[width=16cm]{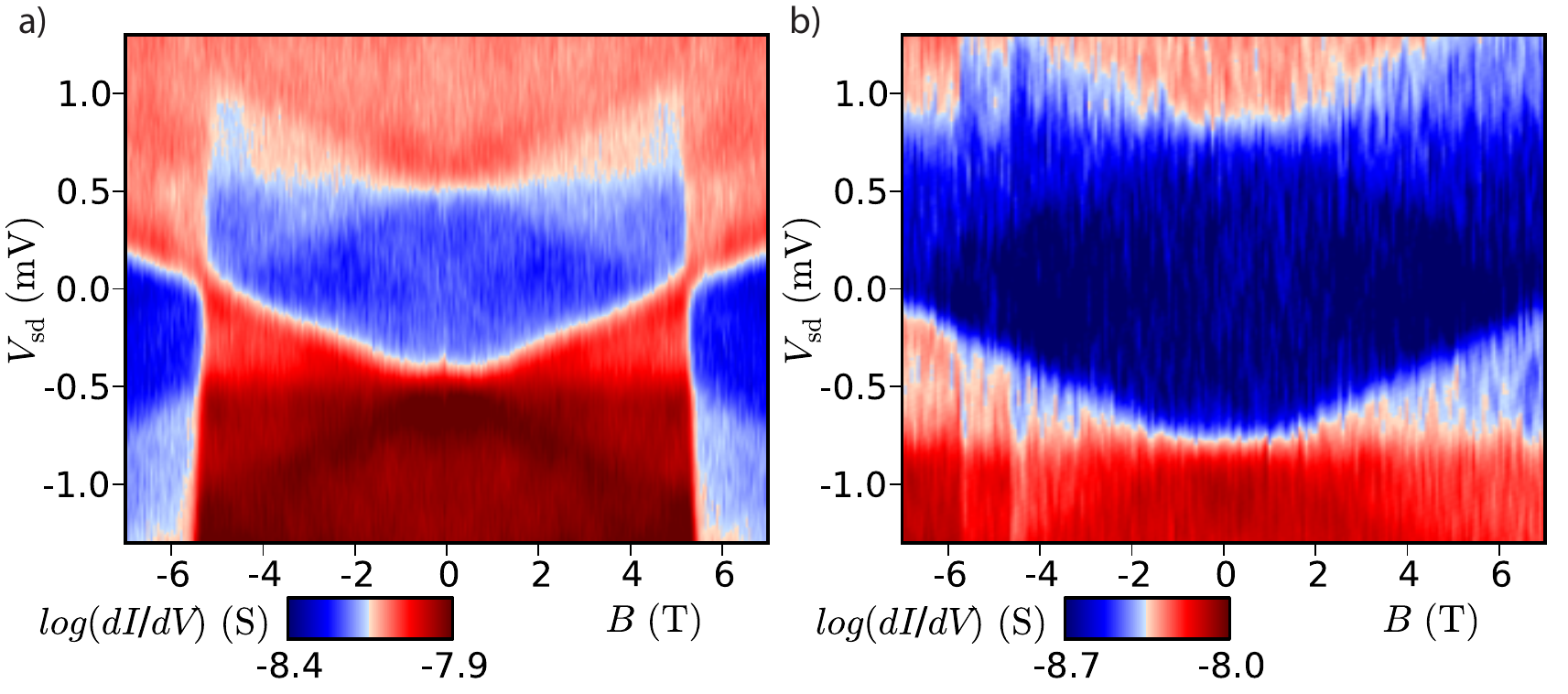}
\caption{Experimental 
differential conductance as a function of magnetic field for a fixed
gate
voltage $V_{\rm{g}} < 
V_{\rm{g}}^{\rm{D}}$ ($n_{\rm{C_{60}}}=1$) for {\bf{(a)}} $J=-0.3$~meV 
 (second run) and {\bf{(b)}} $J=-0.4$~meV (first run). }
\label{figure2}
\end{figure}

In this section, we compare the magnetic field spectroscopy performed
in the 
cotunneling regime, where $n_{\rm{C_{60}}}=1$ ($V_{\rm{g}} < 
V_{\rm{g}}^{\rm{D}}$), for the two runs of 
measurements. As depicted in Figure $4$ of 
the Letter, when the magnetic field increases, the different
multiplet
states are Zeeman split. For a magnetic field $B<B_{\rm{c}}$, the ground 
sate is $|1,+1\rangle$ and taking into account the selection rules, we
observe four
different transitions,
until the 
magnetic field reaches the critical magnetic field $B_{\rm{c}}$ for which
the ground state becomes $|2,+2\rangle$. 
Only two transitions are then allowed, 
obeying $\left|\Delta S^{z}\right| = 0 \; \rm{or} \; 1$, as observed 
in Fig.S.\ref{figure2}(a).
Unfortunately, if the 
critical field $B_{\rm{c}}$ was experimentally accessible 
for the second run corresponding to $J=-0.3$~meV 
(Fig.S.\ref{figure2}(a)), no spin transition 
could be observed for the first run (Fig.S.\ref{figure2}(b)) as the value of
$J=-0.4$~meV was too high. Indeed, as $J$ depends exponentially on the distance 
between the nitrogen and the $\rm{C_{60}}$ cage,
a small deformation can modify $J$
significantly. More precisely, $J$ is
given by an exchange integral, which depends strongly on the electronic
wave function of the $\rm{C_{60}}$ LUMO. This wave function is not only changed by
geometric deformation but also by electrostatics. For example, image
charges in the metal electrodes draw it outwards towards the electrodes
and away from the nitrogen. A modification of the environment can 
thus also lead to a variation of $J$.

\def\figurename{Fig.S.}

\section{Magnetic field spectroscopy in the cotunneling regime for 
$n_{\rm{C_{60}}}=2$}

\begin{figure}
\includegraphics[width=16cm]{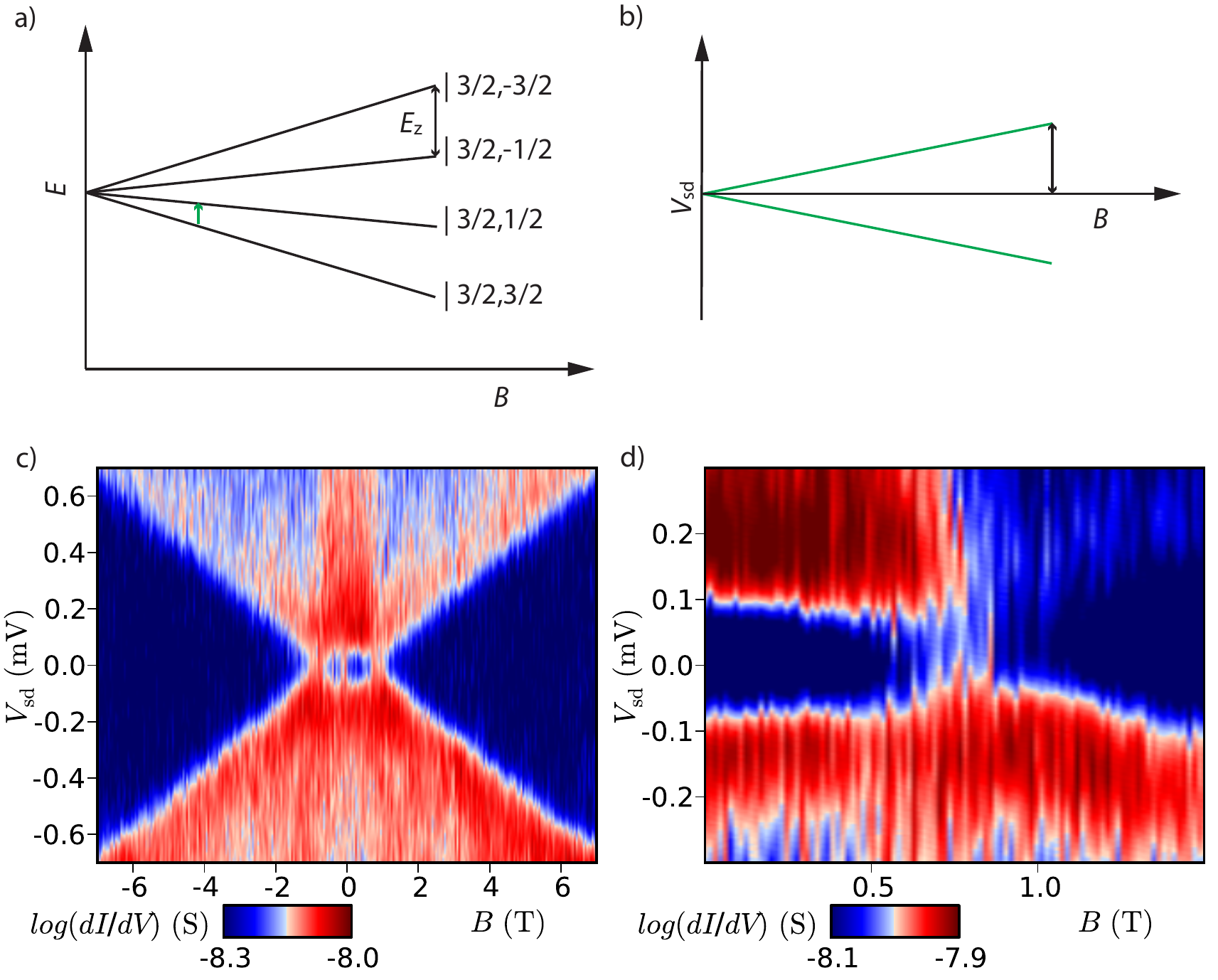}
\caption{ {\bf{(a)}} Zeeman diagram of the multiplet $S = 3/2$ for 
$\rm{N@C_{60}^{2-}}$ ($n_\mathrm{C_{60}}=2$). {\bf{(b)}} Schematic showing the expected  
cotunneling steps deduced
from the Zeeman diagram. {\bf{(c)}} and {\bf{(d)}} $dI/dV$ 
measurements as a function of magnetic field for a fixed gate voltage.}
\label{figure3}
\end{figure}

In the Letter, we raised the question of the unexpected cotunneling steps
at low voltage bias, obtained at zero magnetic field in the 
$n_{\rm{C_{60}}}=2$ region where the total spin of $\rm{N@C_{60}^{2-}}$ is 
$S=3/2$. In Eq. (1) of the Letter, we assume no anisotropy. In this case, 
performing a magnetic field spectroscopy, we should observe 
one cotunneling step increasing linearly with magnetic field 
(Fig.S.\ref{figure3}(b))
corresponding to the transition from the  $|3/2,+3/2\rangle$
state to the $|3/2,+1/2\rangle$ 
state as shown in  Fig.S.\ref{figure3}(a). In
Fig.S.\ref{figure3}(c), we
present the $dI/dV$ measurement as a 
function of magnetic field. We clearly observe that the
cotunneling 
steps evolve linearly in the high magnetic field limit.

\def\figurename{Fig.S.}

However, when we focus on the low magnetic field spectroscopy 
measurements (Fig.S.\ref{figure3}(d)), we observe a non-linear
decrease of the
cotunneling step as a function of magnetic field. This cotunneling 
step and the non linear evolution could be interpreted as a signature 
of anisotropy~\cite{Jo2006,Heersche2006,Zyazin2010,Romeike2003,Timm2007} in $\rm{N@C_{60}^{2-}}$.
This interpretation relies on the assumption of significantly enhanced
spin-orbit coupling on the
nitrogen atom, which might be caused by the hybridization with the
gold leads, as 
the anisotropy of $\rm{N@C_{60}}$ should not exceed tens of 
mT~\cite{Lips2000,Franco2006,Naydenov2006}.

The effect of anisotropy can be included in our model adding
$E_\mathrm{ani} = -K(S^{z})^2$ in our 
Hamiltonian (Eq. 1 of the Letter),
where $E_\mathrm{ani}$ is the anisotropic energy, $K$ is the 
uniaxial anisotropy
constant, and $S_z$ is
the z component of the total spin. However, by including this term 
into
the Hamiltonian, 
we should also observe extra cotunneling steps in the $n_{\rm{C_{60}}}=1$ 
region. We did not measured this signature, either because of a lack 
of sensitivity of our electronic, or because the finite anisotropy 
might be charge dependent, but up to now, we do not have a clear
explanation of the possible charging-state dependence of $K$.